\documentclass{article}    

\usepackage{graphicx}

\title{\textbf{The Parallel Principle}}  
\author{Richard Mould\footnote{Department of Physics and Astronomy, State University of New York, Stony Brook,
\mbox{New York} 11794-3800; http://nuclear.physics.sunysb.edu/ \~{}mould}}  
\date{}    
   
\begin{document}             

\maketitle              

\begin{abstract}

      	Von Neumann's psycho-physical parallelism requires the existence of an interaction between subjective
experiences and material systems.  A hypothesis is proposed that amends physics in a way that connects subjective
states with physical states, and a general model of the interaction is provided.  A specific example shows how the
theory applies to pain consciousness.  The implications concerning quantum mechanical state creation and reduction
are discussed, and some mechanisms are suggested to seed the process.   An experiment that tests the hypothesis is
described elsewhere.  

\end{abstract}

\section*{Introduction}

	I assume that there is a rough correspondence, or at least a working relationship, between the subjective life of
any creature and the objective world in which it is a part.  John von Neumann calls this a \emph{psycho-physical
parallelism}, according to which the images of the creature's psychic experience mirror the objects in its physical
environment \cite{vN}.  Presumably, creature learning begins in infancy by creating a parallelism of this kind that
is practical and useful for the adult.  For this to happen, a conscious species must develop a rudimentary
psycho-physical parallelism at an early stage of evolution.  

	By \emph{consciousness} I mean all that which is contained in the subjective or psychic life of an individual. 
Consciousness is different from the physiological state that gives rise to it, for although it is a by-product of
the physical processes, it is not itself a physical entity.  It is the psycho part of the psycho-physical
parallelism.  

 	Consciousness is widely believed to be \emph{epiphenomenal}, which means that it is created and choreographed by a
physical body but cannot, conversely, influence the behavior of that body.  If that were the case in any species
including our own, then there would be no point to a psycho-physical parallelism.  It would not then matter if a
creature's psychic life mirrored or failed to mirror its physical environment, for having no influence, its psychic
life would not matter to anything at all.  

	The two fundamental disciplines of modern physics, quantum mechanics and general relativity, are mechanically
autonomous.  They provide no mechanism that would allow consciousness to influence the behavior of a physical body;
and accordingly, consciousness can only appear scientifically as an epiphenomenon.  Therefore, barring the
acceptance of the miraculous principle of Pre-Established Harmony proclaimed by Leibnitz, there is no reason to
believe that the subjective life of a conscious being would in any way reflect the physical world in which it
lives.  There is not even reason to believe that the subjective life of a conscious being would be rational; and
certainly, the appearance of rational thinking that parallels the objective world would be enormously improbable. 
So given the present scientific understanding, a psycho-physical parallelism would exist only if there were an
amazing and inexplicable harmony in nature of the kind suggested by Leibnitz.

	I do not accept `pre-established harmony'.  I believe that subjectivity arises naturally within the objective world
in a way that results in a psycho-physical parallelism, and that we can, at least partially, document the reasons
for that development.   To do so, we will have to amend to physics.  

\section*{The Parallel Principle}

	The following statement of the \emph{parallel principle} asserts how subjectivity is generally related to
physiology in humans, and presumably in all conscious species.  

\vspace{0.3 cm}

\emph{The subjective images and ideas of a conscious species are related to its physiology in such a way as to allow
the development of a working psycho-physical parallelism at every stage of evolution.}  

\vspace{0.3 cm}

For this parallelism to work, there must be some degree of mutual monitoring between the psychological and
physiological worlds to keep them together on parallel tracks.  This means that subjectivity must feed information
back to the underlying physiological system, correcting it on the evolutionary stage when it does not create
appropriate (i.e., parallel) images and ideas.  Physiology must respond to this instruction.  

The idea that mind and body must have evolved interactively was discussed by William James, who believed that the
evolution of ``appropriate" subjective feelings would be incomprehensible if feelings were biologically redundant
\cite{WJ}.

On this model, our effort should initially focus on how this parallelism develops in primitive species. 
Our strategy will be to consider amendments to physics that will satisfy the parallel principle in early organisms. 
Attention goes first to the ways in which a creature that is fully automated might begin to experience
consciousness.  

An automaton operates on the basis of a simple stimulus/response sequence, where the success of a sequence is
awarded to the survivor of the evolutionary struggle.  Suppose, as a result of mutation, an amended sequence appears
in the form stimulus/consciousness/response.  The conscious experience in this sequence does not have to be the sole
determinant of the response, but we will allow that it is influential; that is, that it will increase or decrease
the likelihood of one response or another.  If the response favored by the newly introduced consciousness is wrong
(i.e., if it encourages an unfortunate response), then the species will not survive; but if the favored response is
right, then the species will survive.  In the end, a successful species will have a specific conscious experience
that is associated with a successful stimulus/response sequence, and this is the signature of a psycho-physical
parallelism.  A more accurate formulation of the parallel principle might be:

\vspace{0.3 cm}

2nd Formulation: \emph{If an element of consciousness becomes associated with a stimulus/response sequence in a
species, and if it contributes to the long-term survival of the species by enhancing or repressing a response, then
the species will have acquired a rudimentary psycho-physical parallelism.}

\vspace{0.3 cm}

Again, this is because a subjective state that enhances or represses a response will enable the creature to learn
(through evolution) to couple that `psychological' state with a successful `physiological' response.  It is my
belief that the psycho-physical parallelism that we identify with humans began in this way. 

It must be emphasized that the conscious element in the above statement is not just a circuitous way of talking
about another (equivalent) physiological configuration that is itself a response to the stimulation and a
determinant of the response.  That would defeat our purpose by reestablishing an epiphenomenal interpretation; for
again, the content of the conscious element would then be irrelevant to any behavior.  We say, rather, that it is
the conscious element itself that is associated with an enhanced or repressed response.  That is not to say that
consciousness has an existence independent of physiology; for indeed, we assume that it arises out of physiology. 
But we claim that the qualitative properties of the experience are directly related to the enhancement or repression
of a response, and that that correspondence cannot be explained by contemporary physics.  One way of providing the
required feedback is described in the final sections on physics.  

\section*{The General Model \& Hypothesis}

A model is shown graphically in fig. 1 in which a stimulus gives rise to two possible responses $R$  and $R'$,
together with a possible subjectve experience \textbf{E}. This appears in two possible sequences, one
being either $R(\textbf{E})$ or $R'(\emph{nE})$, and another being either $R(\emph{nE})$ or $R'(\textbf{E})$, where
\emph{nE} indicates no associated experience.

\begin{figure}[h]
\centering
\includegraphics[scale=0.8]{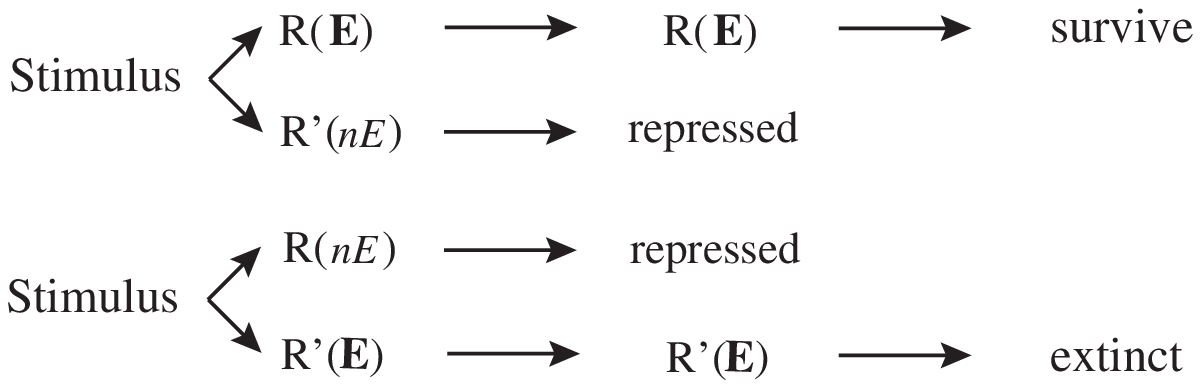}
\center{Figure 1}
\end{figure}

We now add a hypothesis concerning the experience {\bf E} that, we will say in this example, enhances a response.  To
put this hypothesis into play, it will be necessary to amend the underlying physics in a way that will be explained
in a later section.

\vspace{0.3 cm}

The General Hypothesis applied to an enhancing experience:

\emph{When the experience \textbf{E} appears in association with a response, it will ``enhance" the response by
increasing its probability relative to other responses.}  Since the bracketed term \emph{nE} means that there is no
associated experience, it produces no special effect.  

\vspace{0.3 cm}

The claimed influence of \textbf{E} increases the probability of the response $R(\textbf{E})$ in the first
sequence in fig. 1, and $R'(\textbf{E})$ in the second sequence.  Because of normalization, this effectively
represses the responses $R'(\emph{nE})$ and $R(\emph{nE})$.  We make the further assumption that the response $R$ is
favored in the evolutionary struggle, and $R'$ is not.  Therefore, the only species that survives is one in which
the experience \textbf{E} is associated with a life affirming response $R$.  

It should be noted that the experiential mutation introducing \textbf{E} is capable of advancing the evolution of
the species in this example, and it does so without the help of a mechanical mutation of the kind that advances an
automaton.  That is, the sequence in fig. 1 might proceed as indicated without a mechanical mutation also taking
place.  This does not mean that the evolutionary process is thereafter dominated by experiential mutations in
preference to mechanical ones.  However, it is possible that these two processes (mechanical and experiential)
operate independent of one another; and if that is so, they might also operate in tandem.  If that were historically
so, then whatever the relative frequency of the two process, the species would evolve faster than it would with
either one of these processes working by itself.  We would then be able to say that if consciousness is introduced
in a way that gives rise to a psycho-physical parallelism, it will always benefit evolution by increasing the speed
with which a species adapts to its environment.

\section*{A fanciful Example}

An example that I use in previous papers is more specific [3,4].  It involves the experience of ``pain" that is
assumed to \emph{decrease} the probability of any response to which it is associated.  In the interest of
concreteness, a fictitious encounter is imagined between an ancient fish that is initially assumed to be an
automaton, and an electric probe that somehow exists in primordial waters.  The probe provides the stimulus that
gives rise to two possible responses of the fish (1) W-withdraw, or (2) C-continued contact. 

A mutation is assumed to introduce the conscious experience of pain associated with one or the other of these
responses.  The sequence W(no pain) or C(pain) therefore presents itself as a possibility together with the sequence
W(pain) or C(no pain).  In the first case, C(pain) is repressed inasmuch as we require that pain \emph{always}
represses the response with which it occurs.  This leaves a painless withdrawal W(no pain) that will survive the
evolutionary struggle inasmuch as it is a healthy response for the fish.  In the second sequence, W(pain) is
repressed, leaving the fish in painless contact C(no pain) with the probe; and this leads to the fish's demise
inasmuch as that response is unhealthy.  The result is the emergence of a species of fish that instinctively
withdraws from a probe, and at the same time, experiences a release from pain.  We therefore see the beginnings of a
psycho-physical parallelism in which pain is coupled with a dangerous behavior.  

When I speak of ``pain" in this example I do not necessarily refer to the painful experience known to humans. 
Different creatures might experience pain differently.  What is important about pain is the way that it is
associated with the repression of unhealthy of responses.  

I have been alluding to the causal efficacy of consciousness by referring to its ability to `enhance' or `repress'
responses.   I will continue to do so in the interest of simplifying and unifying the discussion.  However, strictly
speaking, one should only talk about possible ``correspondences" or ``associations" between conscious experiences and
physiological responses (or our models of physiological responses).  That's because we can only hope to discover
empirical relationships at this point.  We have no general theory that can explain the psycho-physical interaction
proposed here, and we may never have such a theory.  It is even possible that there is a third unknown (and
unknowable) cause that is common to these relationships [5].  I will nonetheless continue to speak of consciousness
as a `causal' influence because that is the most heuristically effective way of presenting this model.

\section*{The Physics}

A stimulus that acts on a biological organism will generally create a quantum mechanical superposition of body
states over a wide range of possible responses\footnote{It is frequently said that macroscopic states cannot be in
quantum mechanical superpositions because they behave like classical mixtures (i.e., like classical statistical
ensembles).  However, quantum mechanical interference terms do exist between these states when they are taken
together with correlated elements in their environment.  The states in this entanglement appear to be a classical
mixture when the environmental variables are integrated out; but Joss and Zeh call it an ``improper mixture"
because, globally considered, it is a bona fide quantum mechanical superposition with distinct probability
amplitudes.  See refs.\ 6 and 7.  It might equally be called an ``improper superposition".} [6, 7].   I call a
superposition of this kind of an
\emph{endogenous} superposition.  It will consist of many competing physiological configurations, each supporting a
distinctive conscious state, and each with a specific quantum mechanical probability of being realized.  The
external stimulus therefore gives rise to an endogenous superposition of states that are capable of supporting
different degrees and qualities of consciousness.  The probability that one of these states is realized is normally
determined by quantum mechanics alone.  However, I add a special hypothesis concerning pain
consciousness.  

\vspace{0.3 cm}

The Psycho-Physical Hypothesis applied to pain:
 
\emph{When pain consciousness is associated with a component of an endogenous superposition, it will repress that 
component relative to other `painless' components.}

\vspace{0.3 cm}

If more that one component contains pain consciousness, than the degree of repression of each component will be a
function of the intensity of the pain in each.  This hypothesis is entirely qualitative inasmuch as no data is
available to give us a measure of the degree of repression.  It is further limited to one kind of experience - pain
consciousness.  Presumably, pleasurable experiences are associated with enhanced behaviors; and more sophisticated
experience/behavior interactions are dealt with at later stages of evolution.  

Again, the hypothesis is intended to be an amendment to the fundamental mechanics.  It provides feedback from
conscious states to physiological states that is essential if we believe that there is a naturally occurring
psycho-physical parallelism.  The feedback cannot be thought of a euphemism for a physiological activity that is
`really' the underlying cause of the influence; for barring a Leibnizian miracle, a genuine psycho-physical
interaction is necessary for there to be a parallelism.

\section*{State Reduction}

There is still no general agreement concerning how, why, or exactly when a quantum mechanical wave function
collapses upon measurement.  The why of it will not concern us here, but for the parallel principle to work we must
choose a reduction process that satisfies one important condition: namely, that state reduction (or state collapse)
cannot happen too quickly.  A developing endogenous state must have enough time to grow to macroscopic proportions;
and it must have enough time to mature sufficiently to support consciousness.

This condition is automatically satisfied if we adopt the state reduction ideas of John von Neumann.  Accordingly,
(1) state reduction will not occur \emph{unless} a conscious observer is present and aware of the system.  This
means that an endogenous macroscopic state will not collapse until it has matured sufficiently to support a
conscious observer; that is, until an internal self-observation is possible.  This idea is sometimes said to imply
that consciousness \emph{causes} the collapse of a quantum mechanical state function.  As previously stated, I use
terminology like this myself; but one should be reminded that we are talking about empirical relationships in which
consciousness is only found to be associated with state reduction in a certain way.  With this qualification, I
accept the von Neumann account of state reduction.  Without it, a psycho-physical parallelism would not be
possible.

\section*{Seed Particles}

There remains the question of how an endogenous quantum mechanical super-position can be formed in the first
place.  Henry Stapp proposed that the calcium ions that are needed to release neurotransmitters across a synaptic
junction are possible seed particles for the creation of such a superposition \cite{HS}.  Because of the
Heisenberg uncertainty principle, one of these small ions will grow to many times its ``classical" size during the
time it takes for it to diffuse to the vesicles containing neurotransmitters. The resulting uncertainty as to which
transmitters are released is passed on to the neurological level, and this results in the macroscopic uncertainty
implicit in an endogenous superposition.

Other seed mechanisms are possible.  There are many migratory transmitters that travel significant distances from
their point of origin to receptors in other parts of the body, and these can acquire Heisenberg uncertainties in
position.  They are the steroids and peptides that move throughout the body, carried along by blood or
intercellular fluids.  Many of these are small enough and travel for a long enough time to be significantly
affected by Heisenberg uncertainty.  This means that the time of a molecule/receptor attachment is governed by a
quantum mechanical probability distribution.  This in turn leads to an uncertain receptor response.  To this
extent, migratory transmitters guarantee the existence of a superposition of receptors in different stages of
stimulation.  When the resulting uncertainties in all of the body's receptors are taken together, the result will
be a wide-ranging endogenous superposition of possible body states\footnote{A migratory molecule spreads out
spatially as it moves about, due to its Heisenberg uncertainty of momentum.  Its components interact strongly with
the fluids in which it is immersed, so they are also dispersed by such classical mechanisms as diffusion,
turbulence and laminar flow.  Either way, the probability with which a \emph{given} molecule attaches to a
\emph{given} receptor is governed by quantum mechanical uncertainty .  The resulting ensemble of receptor states is
an entanglement of seed molecules and the liquid environment in which they are immersed.  But when that environment
is integrated out, thereby eliminating it as part of the local (macroscopic) system, the resulting receptor states
are found to lack the ability to interfere with one another (see ref. 6, 7).}.

\section*{The Case of Pain}

The \emph{endorphins} produced by the body are migratory molecules that mediate pain by seeking out and attaching to
\emph{opiate receptors} in the brain and other parts of the body.  These molecules are peptides that are small
enough and generally travel far enough to seed endogenous quantum mechanical superpositions that include a broad
range of states with different degrees of pain consciousness.  Endorphins can therefore function as the pain
suppressers in the fanciful example of the fish, and this gives them a possible evolutionary role of some
importance.  Opiate receptors have been found in very ancient species, going back to the early vertebrates
\cite{CP}.  Since they serve no other purpose than to stimulate analgesic and euphoric effects, and since they have
such a pervasive presence in all vertebrates, it is easy to believe that opiate receptors served a compelling
evolutionary purpose associated with the most elementary of conscious experiences\footnote{In advanced species,
receptors perform these functions as well as modify more sophisticated moods in the direction of analgesia and
euphoria. See ref. 10.}$^,$\footnote{The existence of ancient opiate receptors does not in itself prove the
existence of consciousness, inasmuch as an automaton might very well use these devices to modify a response to
certain kinds of stimuli.  However, I believe that consciousness was introduced through these devices.  The extent
to which they existed before that event, or came into existence as party to that event, is not a question that I
address here.} 
\cite{CL}.   

Because of its (Heisenberg) uncertainty of position, a migrating endorphin molecule has a less-than-one probability
of attaching itself to an opiate receptor at any given moment.  So the total number of receptors that are turned on
at that moment is quantum mechanically uncertain.  This number is a variable of a quantum mechanical superposition,
each component of which potentially supports different degrees of pain consciousness.  Our psycho-physical
hypothesis tells us that those components with a greater degree of pain will be repressed relative to those with a
lesser degree of pain; and as a result, the distribution of states in the superposition will shift in the direction
of states with lesser pain.  The probability that the subject will experience less pain is thereby increased, thus
establishing a connection between subjective experience and physiology as required by the psycho-physical
parallelism.  

It is difficult to imagine how this theory can be tested if the only seed particles available are the calcium ions
within neuron synapses.  But small migrating molecules are a different story.  Exogenous opiates, such as codeine,
morphine, heroin also alleviate pain and/or give pleasure by attaching to opiate receptors \cite{SS}. These
molecules are small enough and they travel far enough to be seed particles, and they are easier to manipulate
experimentally.  It is therefore possible to test the above theory by injecting pharmacological doses of these
opiates into subjects, determining the extent to which they attach to receptors in conscious subjects; and
comparing this with their attachment in subjects who receive subpharmacological doses.  The author has proposed an
experiment along these lines that injects synthetic opiates into humans using positron emission tomography (PET),
or into rats using autoradiography \cite{RM3}.

\end{document}